# Experimental investigation of solubility trapping in 3D printed micromodels

Alexandros Patsoukis Dimou[1,3] · Mahdi Mansouri Boroujeni [2] · Sophie Roman[2] · Hannah P. Menke[1] · Julien Maes[1]

[1]Institute of Geoenergy and Engineering, Heriot-Watt University, Edinburgh, UK
[2]Institut des Sciences de la Terre d'Orléans, Univ. Orléans, CNRS, BRGM, ISTO UMR 7327, F–45071 Orléans, France
[3]Instite of Fluid Science, Tohoku University, 2 Chome-1-1 Katahira, Aoba Ward, Sendai, Miyagi 980-8577, Japan

## Abstract

*Hypothesis:* Understanding interfacial mass transfer during dissolution of gas in a liquid is vital for optimising large-scale carbon capture and storage operations. While the dissolution of $CO_2$ bubbles in reservoir brine is a crucial mechanism towards safe $CO_2$ storage, it is a process that occurs at the pore-scale and is not yet fully understood. Direct numerical simulation (DNS) models describing this type of dissolution exist and have been validated with semi-analytical models on simple cases like a rising bubble in a liquid column. However, DNS models have not been experimentally validated for more complicated scenarios such as dissolution of trapped $CO_2$ bubbles in pore geometries where there are few experimental datasets. In this work we present an experimental and numerical study of trapping and dissolution of $CO_2$ bubbles in 3D printed micromodel geometries

*Experiments:* We use 3D printing technology to generate three different geometries, a single cavity geometry, a triple cavity geometry and a multiple channel geometry. In order to investigate the repeatability of the trapping and dissolution experimental results, each geometry is printed three times and three identical experiments are performed for each geometry. The experiments are performed at low capillary number (Ca=$3.33 \times 10^{-6}$) representative of flow during $CO_2$ storage applications. DNS simulations are then performed and compared with the experimental results.

*Findings:* Our results show experimental reproducibility and consistency in terms of $CO_2$ trapping and the $CO_2$ dissolution process. At such low capillary number, our numerical simulator cannot model the process accurately due to parasitic currents and the strong time-step constraints associated with capillary waves. However, we show that, for the single and triple cavity geometry, the interfacial transfer and resulting bubble dissolution can be reproduced by a numerical strategy where the interfacial tension is divided by 100 to relax the capillary time-step constraints. The full experimental dataset is provided and can be used to benchmark and improve future numerical models.

## 1. Introduction

Accurate prediction of gas dissolution during flow in porous media is vital for $CO_2$ storage applications [1, 2]. During $CO_2$ storage, large quantities of the captured $CO_2$ are injected in the pore-space of an underground reservoir (e.g., saline aquifer, depleted oil and gas field) for permanent storage. The trapping can then occurs with four distinct mechanisms: (1) structural trapping, where an impermeable cap rock restricts the $CO_2$ to escape the formation rock, (2) residual trapping, where the $CO_2$ is trapped in disconnected clusters inside the reservoir rock, (3) solubility trapping, where the $CO_2$ is dissolved inside the formation water and (4) mineral trapping, which refers to the mineralization of $CO_2$ via reactions with minerals, aqueous phase and organic matter inside the formation [3].

While we are mainly interested in the accurate prediction of the storage at the reservoir scale [4-6], three of the four trapping mechanisms (residual, solubility, and mineral) primarily occur at the pore-scale. It is therefore crucial to accurately integrate these pore-scale mechanisms into reservoir-scale models, a process referred to as upscaling. Although upscaling of residual trapping at the pore-scale has been extensively investigated [7-9], these investigations seldom include the effect of solubility trapping, mainly because numerical models of multiphase flow at the pore-scale that include interfacial transfer are not ready for simulations in micro-CT images at reservoir conditions.

Computational Fluid Dynamics (CFD) are an essential tool that complement experiments while allowing for sensitivity analysis of various physical parameters. Numerical simulations describing two-phase flow can be performed using the algebraic Volume-of-Fluid (VOF) method [10]. In the VOF method, the interface between the two fluids is captured using an indicator function. Interface transfer can be modelled with the VOF method using a single-field approach and the Continuous Species Transfer (CST) method [11, 12]. This method has been extended to simulate the local volume change resulting from the interfacial transfer, and used to successfully capture dissolution of a rising gas bubble in a liquid column [13]. However, there is lack of an experimental dataset that allows benchmarking of the existing model during complicated scenarios where the fluid/solid interaction impacts the dissolution process, such as trapped $CO_2$ bubbles in a pore-space. Simulations of the dissolution of a single $CO_2$ bubble trapped in a single cavity have been performed [14, 15], but have not been experimentally validated. Multiphase flow in porous media is characterised by the capillary number, which is the ratio of viscous forces to capillary forces. Because the VOF suffers from the well-known problem of parasitic currents at low capillary numbers ($Ca < 10^{-5}$) [16] where capillary forces are dominant, the capability of the model to simulate gas bubbles dissolution at flow rates representative of $CO_2$ geological storage is unclear.

Micromodels are artificial replicas of natural porous media with standardised, repeatable geometries and have contributed to our understanding of pore-scale physics [17-22]. Micromodel experiments are advantageous because experiments can be conducted in the same geometry with multiple experimental protocols. Furthermore, they allow for geometry control of porous media and therefore enable the precise investigation of the impact of pore structure on flow. Due to their transparent nature, micromodels also permit direct fluid flow observation with a high-speed, high-resolution camera. Micromodels have thus been widely utilized to investigate physio-chemical phenomena in geoscience. Sun and Cubaud [23] used single channel microfluidics to investigate $CO_2$ bubble dissolution in water, ethanol, and methanol inside a channel. Amarasinghe, Farzaneh [24] used a microfluidic device to visualise convective mixing of $CO_2$ in water and n-decane. Furthermore, pore-scale $scCO_2$ dissolution experiments have been conducted in heterogeneous micromodels for both imbibition and drainage processes [25-28]. However, micromodel fabrication techniques like etching and moulding are slow, expensive, and limited in 2D and 2.5D structures. While microfabrication of micromodels with polydimethylsiloxane (PDMS) is inexpensive and allows for investigation of fluid flow, PDMS is permeable to gas [29, 30] and thus not suitable for studying $CO_2$ dissolution. An alternative to micromodels made of PDMS is micromodels made of glass. However, fabrication of micromodels made of glass is expensive and therefore investigating the impact of small geometrical alterations to the over flow would be commercially difficult.

The emergence of additive manufacturing (AM), also called 3D printing, allows for a compelling alternative to conventional micromodels. 3D printing converts computer-assisted design (CAD) into a physical object in a single additive process. Commercial 3D printers, which can produce structures ranging from a few microns to several centimetres, are beginning to challenge conventional microfluidic fabrication techniques as the research prototyping approach to micro-fabrication (Gonzalez et al., 2022). 3D printing has been applied to a wide range of industries including medicine [31], biomedical engineering [32] and aerospace engineering [33]. Compared with standard micromodel fabrication techniques, the attraction of 3D printing is twofold. First, 3D printing has the potential to fabricate in three dimensions in a way that was not previously possible. Secondly, the inexpensive nature of the 3D printed micromodels combined with the fast fabrication (~3 h) allows experimental investigations in multiple geometries quickly, allowing for optimisation of the geometry that will produce the best experimental dataset [34]. Small alterations of 3D printed models also enable geometrical sensitivity analysis, something that is not commercially possible with micromodels made of glass. Two-dimensional single-layer 3D printed micromodels have already attracted attention and have been used to investigate pore-scale phenomena relevant to flow during subsurface processes [35-39]. 3D printed micromodels have also been experimentally validated as devices to conduct fluid flow experiments [40] and have been used for water-$CO_2$ multiphase flow experiments [41].

In this paper we have two objectives: (1) to investigate whether we can generate identical 3D printed microfluidic devices where trapping and dissolution of $CO_2$ can occur repeatably and generate a benchmark dataset; (2) to investigate whether existing direct numerical simulations describing $CO_2$ dissolution can accurately capture the dissolution rate obtained from the experiments conducted in the 3D printed micromodels. The method for the micromodel generation is presented in Section 2. We then introduce the experimental apparatus in Section 3 and the image processing in Section 4. We introduce the numerical methods in Section 5 and present the results and conclusions of this work in Sections 6 and 7.

## 2. Micromodel Generation

The 3D printed micromodel devices were designed with the use of the OpenSCAD open-source software (https://openscad.org/index.html), which meshes the structure into the stl format. Three identical microfluidic devices were printed for each of the three different geometries. The first geometry consists of a single cavity of 1 mm width and 2 mm height connected with single channel of 1 mm width (SC) (Figure 1(A)). The second microfluidic design consists of 3 cavities of 1mm width and 2 mm height. The cavities are placed at a 2 mm distance from each other and connected with a single channel with 1 mm width (TC) (Figure 1(B)). The third design consists of a top 4 mm channel connected perpendicularly via five 500 μm channels to a 500 μm bottom channel (MC). (Figure 1(C)). The overall depth of the patterns was set to 1 mm for all micromodel designs.

The printer used for the creation of the micromodels in this project is the Formlabs Form 2 stereolithography (SLA). Formlabs Form 2 works by successively solidifying layers of liquid photopolymer resin one on top of the other. A detailed description on the printing process and the resolution of the printer can be found in Patsoukis Dimou, Menke [40].

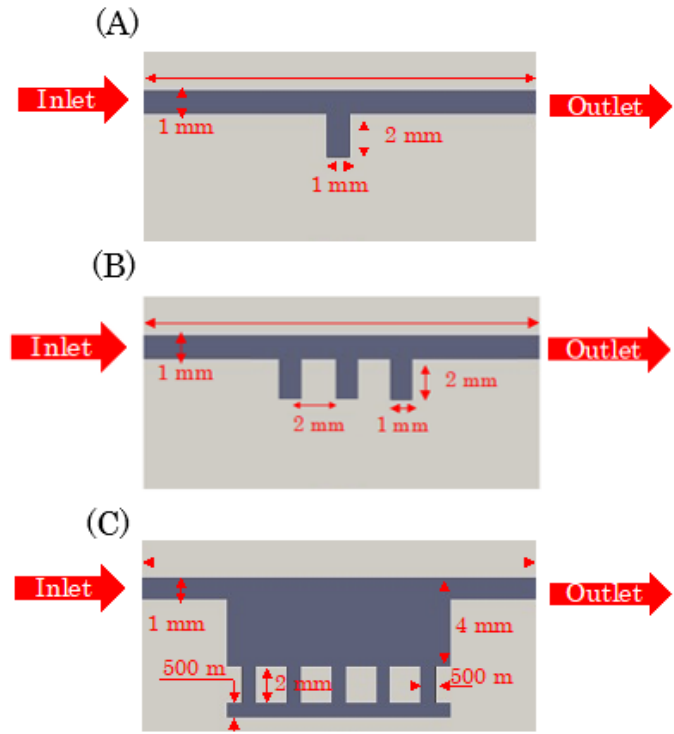

Figure 1: Micromodel geometries generated for conducting $CO_2$ dissolution experiments. Single Cavity design (SC) (A), Triple Cavity design (TC) (B), Multiple Channels design (MC) (C). Gray indicates solid walls and dark blue indicates the fluid path.

## 3. Experimental Setup

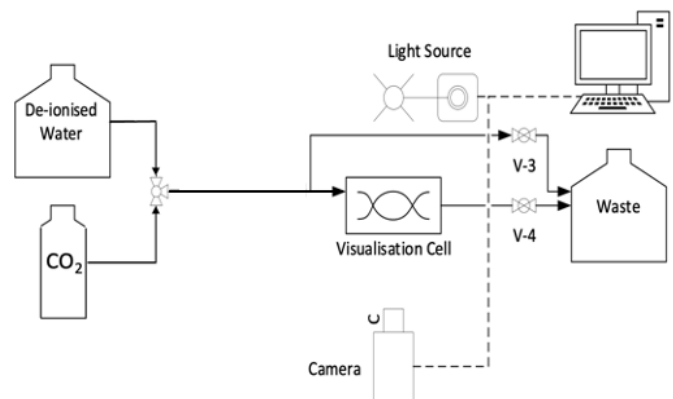

Figure 2: Experimental setup used to investigate dissolution of the $CO_2$. Syringe Pump (Chemyx Fusion 4000), Light source (SCHOTT COLD Vision Light Source), $CO_2$ Cartridge, Pressure regulator (Pera Keg Charger), High resolution camera (Baumer VCXU 51), Perspex Visualisation Cell.

Once the micromodel was successfully printed, it was inserted in the perspex transparent visualisation cell face-down and sealed using an o-ring. 1/16-inch peek tubing was used to connect the syringe pump (Chemyx Fusion 4000) and a pressure regulator (Pera Keg Charger) to a three-way valve and 1/16-inch peek tubing to the bypass, visualisation cell, and then to the outlet. A Baumer VCXU 51 high-resolution camera which allowed for a 3.45 μm/pixel resolution was mounted beneath the flow cell and recorded images using the Stream Pix 11 software (https://www.norpix.com/products/streampix). Above the visualisation cell, a LED light source (SCHOTT ColdVision Light Source) was installed to reduce shadows and enable clear visualisation of the fluid movement.

For each experiment, pure $CO_2$ was injected in the system for 10 minutes by imposing a pressure differential of 3.7 Pa with a total of over two hundred pore volumes (PV) injected through the system to ensure full saturation of the micromodel with $CO_2$. Distilled water was then injected inside the micromodel at constant flow rate and the trapping and dissolution of $CO_2$ was recorded with the high-resolution camera at an acquisition speed of 1 frame per second. A flow rate of $Q = 1.67 \times 10^{-10}$ m$^3$.sec$^{-1}$ was used for all experiments.

## 4. Image Processing

Images were post-processed in MATLAB® to improve the measurement quality. First the images were denoised using the contrast limited adaptive histogram equalization [42] and the adaptive wiener denoise filter [43]. The pore-space was then segmented and the water, solid and $CO_2$ phases were separated. Finally, the pixels representing the $CO_2$ phase were counted in every successive image. Due to the inability to observe the curvature of the trapped bubbles in the Z axis, the pixels measured in every successive image represent the fraction of $CO_2$ in the middle plane as demonstrated in Figure 3.

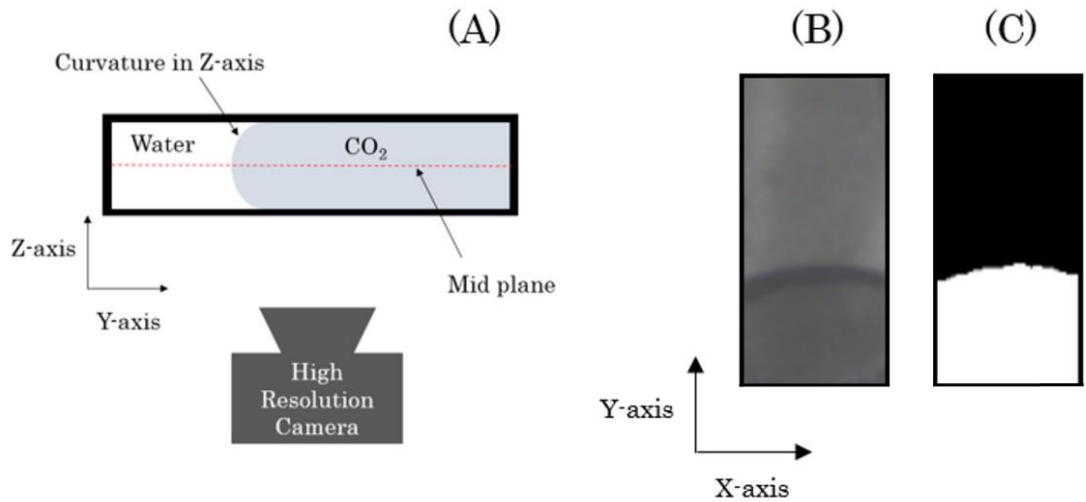

*Figure 3: (A) Conceptual image explaining the inability to observe the curvature of the interface between the water and $CO_2$ phase at the Z axis. Due to this inability, the fraction of $CO_2$ corresponding to the middle plane (red line) is measured. (B) Raw image of cavity. (C) Segmented image of cavity.*

## 5. Numerical Methods

### 5.1 The Volume-Of-Fluid Method

In the volume of fluid method, the location of the interface is given by the indicator function $\alpha$, which is equal to the volume fraction of one phase in each grid cell. The density and viscosity of the fluid are given by volumetric averaging

$$\rho = \rho_1 \alpha + \rho_2(1 - \alpha), \qquad (1)$$

$$\mu = \mu_1 \alpha + \mu_2 (1-\alpha), \tag{2}$$

where $\rho_i$ (kg.m$^{-3}$) and $\mu_i$ (Pa.s) are the density and viscosity of phase $i$. The velocity, pressure and species concentration in the domain are expressed in terms of single-field variables in a similar manner

$$\boldsymbol{u} = \boldsymbol{u}_1 \alpha + \boldsymbol{u}_2 (1-\alpha), \tag{3}$$

$$p = p_1 \alpha + p_2 (1-\alpha), \tag{4}$$

$$c = c_1 \alpha + c_2 (1-\alpha), \tag{5}$$

where $u_i$ (m.s$^{-1}$), $p_i$ (Pa) and $c_i$ (kg.m$^{-3}$) are the velocity, pressure and species concentration in phase $i$. The phase advection equation writes

$$\frac{\partial \alpha}{\partial t} + \nabla \cdot (\alpha \boldsymbol{u}) + \nabla \cdot (\alpha(1-\alpha)\boldsymbol{u}_r) = \frac{\dot{m}}{\rho_1}, \tag{6}$$

where $\dot{m}$ (kg.m$^{-3}$.s$^{-1}$) is the mass transfer from phase 2 to phase 1 and $\boldsymbol{u}_r = \boldsymbol{u}_1 - \boldsymbol{u}_2$ is the relative velocity of the interface between the two phases [44]. To limit numerical diffusion at the two-fluid interface, the relative velocity is modelled as a compressive velocity

$$\boldsymbol{u}_r \equiv \boldsymbol{u}_{comp} = \boldsymbol{n}_\Sigma \left[ min\left( c_\alpha \frac{|\phi_f|}{A_f}, max\left(\frac{|\phi_f|}{A_f}\right) \right) \right], \tag{7}$$

where $c_\alpha$ is the compression constant (generally between 0 and 4), $A_f$ is the area of face $f$ and $\phi_f$ the volumetric flux across $f$. In all our simulations, $c_\alpha=1.0$.

Assuming each phase is Newtonian and incompressible, and neglecting gravity effects as well as assuming the fluid properties are constant in each phase, the single-field momentum equation can be written as [45]:

$$\nabla \cdot \boldsymbol{u} = \dot{m} \left( \frac{1}{\rho_1} - \frac{1}{\rho_2} \right), \tag{8}$$

$$\frac{\partial \rho \boldsymbol{u}}{\partial t} + \nabla \cdot (\rho \boldsymbol{u}\boldsymbol{u}) = -\nabla p + \nabla \cdot \left( \mu (\nabla \boldsymbol{u} + \nabla \boldsymbol{u}^T) \right) + \boldsymbol{f}_{st}, \tag{9}$$

where $\boldsymbol{f}_{st}$ (kg.m$^{-2}$.s$^{-2}$) is the surface tensions force, calculated using the Continuous Surface Force (CSF) method [46],

$$\boldsymbol{f}_{st} = \sigma \kappa \nabla \alpha. \tag{10}$$

$\sigma$ (N.m$^{-1}$) is the interfacial tension, $\kappa$ (m$^{-1}$) is the curvature at the interface which can be calculated as

$$\kappa = -\nabla \cdot \boldsymbol{n}_\Sigma, \tag{11}$$

where $n_\Sigma$ is the interface vector defined as

$$\boldsymbol{n}_\Sigma = \frac{\nabla \alpha}{\|\nabla \alpha\|}. \tag{12}$$

Assuming that the gas phase is pure and that the gas dissolves in the liquid phase with Henry's constant $H$ and remains diluted, the single-field concentration equation satisfies the advection-diffusion equation given by the Continuous Species Transfer (CST) formulation

$$\frac{\partial c}{\partial t} + \nabla \cdot \boldsymbol{F} + \nabla \cdot \boldsymbol{J} = 0, \tag{13}$$

where $\boldsymbol{F}$ (kg.m$^{-2}$.s$^{-1}$) is the advective flux and $J$ (kg.m$^{-2}$.s$^{-1}$) is the diffusive flux. Maes and Soulaine [13] showed that the advective flux can be written as

$$\boldsymbol{F} = c\boldsymbol{u} + \alpha(1-\alpha)\frac{\nabla c \cdot \nabla \alpha}{\|\nabla \alpha\|^2}\boldsymbol{u}_r, \tag{14}$$

and the diffusive flux can be written as [13]

$$\boldsymbol{J} = -D^{SF}\nabla c + \boldsymbol{\Phi}, \tag{15}$$

where $\boldsymbol{\Phi}$ (kg.m$^{-2}$.s$^{-1}$) is the CST flux and $D^{SF}$ (m$^2$.sec$^{-1}$) is the single-field diffusion coefficient. The CST flux can be written as

$$\boldsymbol{\Phi} = (1-\mathrm{H})D^{SF}\frac{c}{\alpha + H(1-\alpha)}\nabla \alpha, \tag{16}$$

and the single-field diffusion coefficient can be expressed as

$$D^{SF} = \frac{\alpha D_1 + H(1-\alpha)D_2}{\alpha + H(1-\alpha)}, \tag{17}$$

where $D_i$ (m$^2$.sec$^{-1}$) is the diffusion coefficient of phase $i$. Finally, the mass transfer $\dot{m}$ at the interface can be calculates as

$$\dot{m} = -\frac{D^{SF}\nabla c - \boldsymbol{\Phi}}{1-\alpha} \cdot \nabla \alpha. \tag{18}$$

The system of equations can be characterized by three dimensionless numbers. These are the Reynolds number,

$$Re = \frac{\rho_1 UL}{\mu_1}, \tag{19}$$

which is the ratio of inertial forces to viscous forces, the Péclet number

$$Pe = \frac{UL}{D_1}, \tag{20}$$

which is the ratio of advection to diffusion, and the Capillary number

$$Ca = \frac{\mu_1 U}{\sigma},\qquad(21)$$

which is the ratio of viscous to capillary forces. In Equations 19, 20 and 21, $U$ is the reference velocity which is the velocity inside the inlet channel for the SC, TC and MC geometry. And $L$ is the characteristic length, which is the width of the inlet channel.

In reality, as $CO_2$ dissolves in the water phase, it dissociates into $H^+$ and $HCO_3^-$. Nevertheless, in our simulation framework we neglect any chemical reactions, and we assume a single component bubble where only one species exists in the $CO_2$ phase, which can be transferred to the water phase. The fluid properties for the water and the $CO_2$ used in this work are summarised in Table 1. The dimensionless numbers $Re$, $Pe$ and $Ca$ for the three geometries and the flow rates are given in Table 2.

*Table 1: Fluid properties for $CO_2$ and water*

| Phase | Density $(Kg.m^{-3})$ | Dynamic Viscosity $(Pa.s)$ | Henry's Constant $(-)$ | Interfacial tension $(mN.m^{-1})$ | Diffusion Coefficient $(m^2.s^{-1})$ |
|---|---|---|---|---|---|
| $CO_2$ | 1.87 | $0.81 \times 10^{-5}$ | 1.25 | | 1.6 |
| Water | 1000 | $10^{-6}$ | 0 | 50 | |

*Table 2: Reynolds, Peclet and Capillary Number for the three micromodel geometries.*

| Model | Re (-) | Pe (-) | Ca (-) |
|---|---|---|---|
| SC | 0.17 | 104.17 | $3.33 \times 10^{-6}$ |
| TC | 0.17 | 104.17 | $3.33 \times 10^{-6}$ |
| MC | 0.17 | 104.17 | $3.33 \times 10^{-6}$ |

The capillary numbers obtained in the experiments, which are as low as $Ca = 3.33 \times 10^{-6}$. DNS of multiphase flow at such low capillary number is challenging for two reasons. The presence of parasitic currents is a well-known challenge [46-50]. Parasitic currents are located close to the fluid-fluid interface and in our case can impact the volume of the $CO_2$ that will be trapped inside the cavity during the displacement, and the dissolution of the gas phase once it has been trapped. Additionally, the propagation of capillary waves imposes a stability time-step constraint known as the capillary wave or Brackbill time-step constraint [46]

$$\Delta t < \Delta t_B = \sqrt{\frac{(\rho_1 + \rho_2)\Delta x^3}{4\pi\sigma}}.\qquad(22)$$

where $\Delta x$ is the grid block size (m). For the fluid properties presented in Table 1, and for a mesh resolution $\Delta x = 50$ μm, we obtain $\Delta t_B = 2.17 \times 10^{-6}$. Therefore, to simulate an injection time of 5 mins would require $1.35 \times 10^8$ time-steps, and the computational time would be too restrictive.

One possible way to simulate the experiments conducted in the SC and the TC geometries is by initializing the bubbles inside the cavities and only simulate the dissolution of the bubbles after they have been trapped with a reduced surface tension of $\sigma/100$. Indeed, as long as the bubbles remain at capillary equilibrium at all times, the value of the interfacial tension does not impact the flow behaviour. To ensure that the bubbles remains at capillary equilibrium, the pressure inside the bubbles is monitored and compared to the capillary pressure obtained using the Young-Laplace equation

$$P_c = \frac{2*\sigma*cos\theta}{r}. \qquad (23)$$

Where $\theta$ is the contact angle (º) and $r$ is the effective radius of the interface (m).

Using this method, a DNS simulation can be performed for the SC and TC geometries. However, a DNS simulation for the MC geometry cannot be conducted. This is because in the MC geometry the dissolution of the $CO_2$ is coupled with displacement of the $CO_2$. Such a process is strongly affected by the surface tension and therefore reducing the surface tension will lead to erroneous results. Furthermore, due to the increased complexity and size of the domain, the mesh size is larger than the rest of the scenarios making the computational time significantly higher.

The numerical method has been implemented in GeoChemFoam (https://github.com/geochemfoam) following the work of Maes and Soulaine [13] . To mesh the computational domain, a 3D uniform cartesian grid was first generated, and then cells containing solid were removed and replaced by cartesian cells to match the solid boundaries using the OpenFOAM snappyHexMesh utility. Given the size of the computational domain, a mesh with a cell resolution of 50x50 μm was used. The flow rates used for the simulations are the same as the flow rates used for the experiments.

Finally, after the numerical simulations have been completed, to achieve accurate comparison with the experiments we extracted and measured the fraction of $CO_2$ that occupies the middle plane inside the cavity of the DNS 3D geometry which is what is measured with the experiments conducted with 3D printed micromodels. This is demonstrated in Figure 4 for the SC geometry.

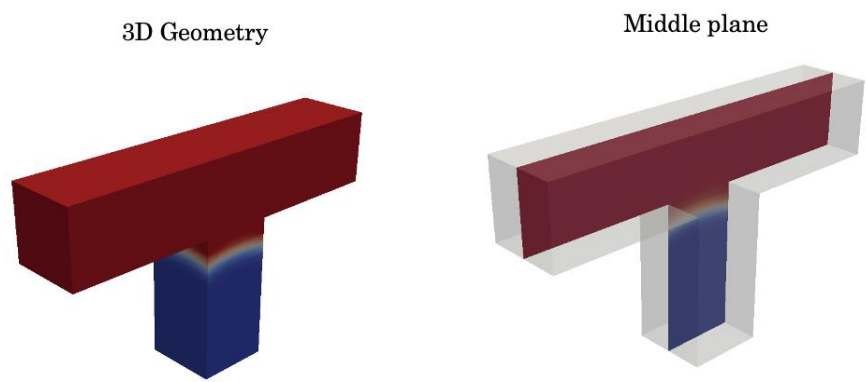

*Figure 4: Schematic of the middle plane of the 3D geometry which is used for direct comparison with the experiments conducted in the 3D printed micromodels.*

# 6. Results & Discussion

Three $CO_2$ trapping and dissolution experiments are performed in various geometries and the repeatability of the trapping and dissolution of the $CO_2$ in each geometry is investigated. DNS simulation of the dissolution of the $CO_2$ is then performed to investigate whether DNS can accurately capture the experimental results.

## 6.1 Single cavity configuration

Three experiments are conducted in three identical SC micromodel geometries. The fraction of $CO_2$ inside the cavity is measured. The area can be seen in red in Figure 5. Until the moment $CO_2$ enters the cavity, it is fully saturated with $CO_2$ and the fraction is unity. In all experiments at $T = 0.5$ min water invades the cavity and traps the $CO_2$. The amount of $CO_2$ trapped inside the cavity is reported in Table 3. The maximum fraction of $CO_2$ trapped inside the cavity is 0.9 while the minimum value trapped is 0.88 leading to a variation of 2% between the experiments. The overall change of the $CO_2$ inside the cavity is shown in Figures 5 and Figure 7. All experiments show the same trend of dissolution rate for the $CO_2$ bubble inside the cavity. The $CO_2$ initially dissolves at a high rate but decreases later as the distance of the water-$CO_2$ interface from the channel increases. These results indicate that initially the dissolution is dominated by the advection caused by the constant flow inside the channel, but as the $CO_2$ dissolves and moves further from the channel the dissolution starts being dominated by diffusion rather than advection. The good agreement between the experimental results conducted in three identical 3D printed micromodels shows that 3D printed single cavity geometry can be generated repeatably and multiphase flow and dissolution experiments can be reproduced. Furthermore, the reproducibility provides confidence that the results could be used to validate direct numerical simulation models.

*Table 3: Fraction of $CO_2$ trapped inside the middle of cavities for the experiments conducted in the SC geometriess.*

| Geometry | Trapped $CO_2$ SC1 | Trapped $CO_2$ SC2 | Trapped $CO_2$ SC3 |
|---|---|---|---|
| SC | 0.89 | 0.90 | 0.88 |

For the DNS benchmark simulation, the bubble is initialised inside the cavity so that the fraction of $CO_2$ is 0.89, matching the bubble size trapped in the experiments. At $T = 0.5$ pure water is injected in the left side of the channel at a flow rate $Q = 1.67 \times 10^{-10}$ m$^3$.sec$^{-1}$, which is $Ca = 3.33 \times 10^{-6}$. The results of the numerical simulation can be seen in Figure 6 and Figure 7. When comparing the numerical simulation results to the experimental results at times $T = 0.5, 4.5$ and $8.5$ min we can observe that the numerical simulation accurately captures the mass evolution of the $CO_2$ bubble. In Figure 7, it can be also seen that the dissolution rate predicted by the numerical simulation matches the dissolution rate obtained from the experimental results until time $T = 6$ min. After this point the DNS results start to diverge from the experimental results. This could be due to several reasons. It could be due to numerical error resulting from parasitic currents, a lack of convergence of the numerical solution and therefore a need to increase the mesh resolution, or a lack of accuracy in the transport parameter $H$ and $D$. To further investigate the reason of deviation would require improvement of the computationally efficiency of the solver and reduction of the large time-step restrictions. Figure 8 shows the capillary pressure at times $T = 0.5, 4.5$ and $8.5$ minutes. The capillary pressure stays constant at a value of $P_c = 1$ Pa, which indicated that the bubble is at capillary equilibrium;

therefore, the surface tension does not impact the dissolution of the CO$_2$ bubble, which is consistent with our assumption.

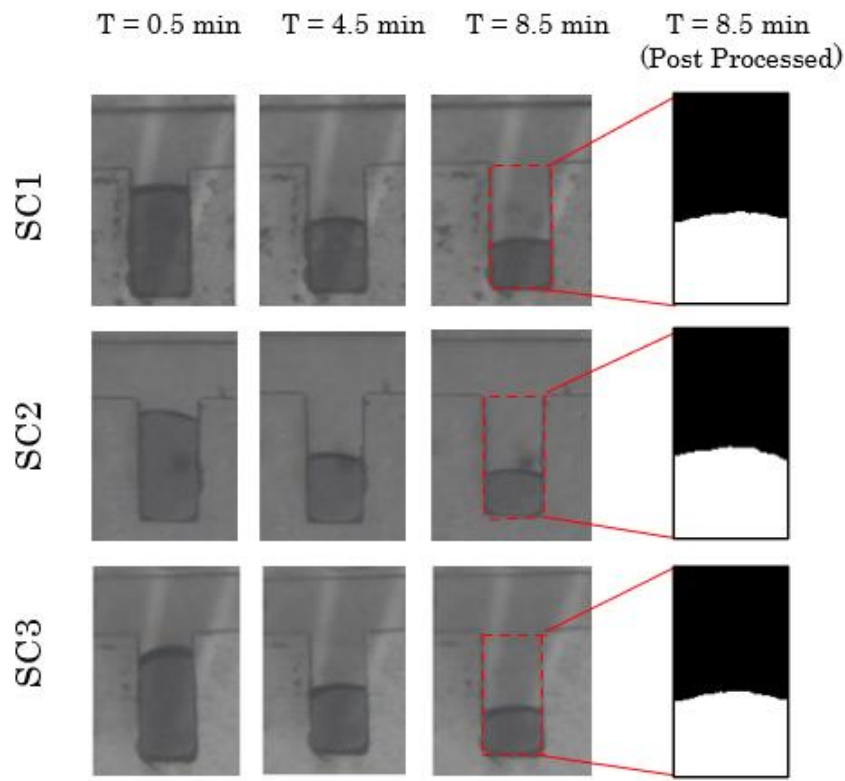

*Figure 5: Dissolution of the CO$_2$ bubble trapped in the cavity. CO$_2$ bubble size observed during the experiment at times T= 0.5, 4.5 and 8.5 minutes.*

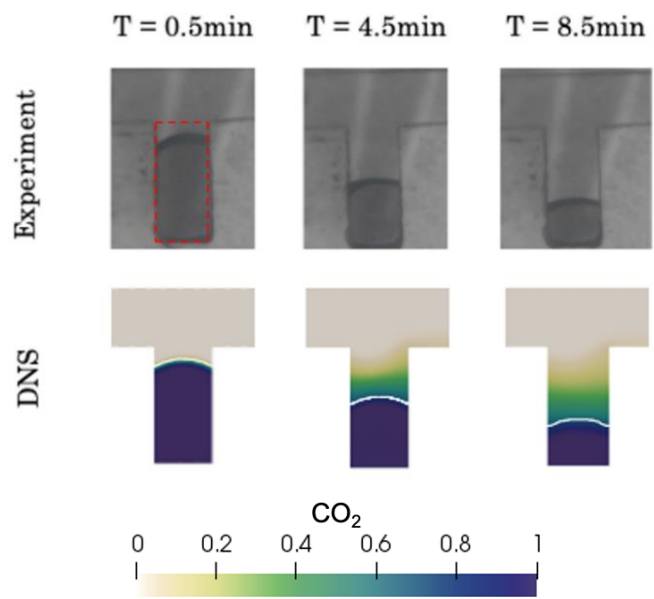

*Figure 6: Dissolution of the CO$_2$ bubble trapped in the SC geometry. CO$_2$ bubble size observed during the experiment at T =0.5, 4.5 and 8.5 min (top). CO$_2$ bubble obtained from the direct numerical simulation at times T = 0.5, 4.5 and 8.5 minutes (bottom).*

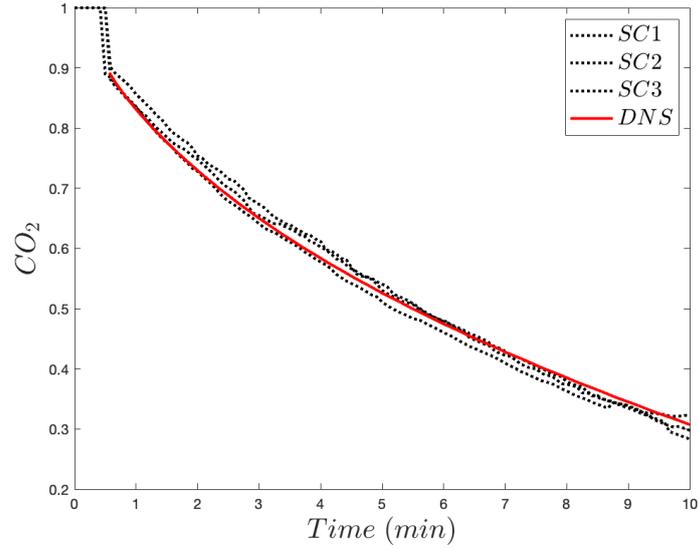

*Figure 7: Fraction of CO₂ inside the cavity, obtained from the experiments conducted in the 3D printed micromodels, and the DNS.*

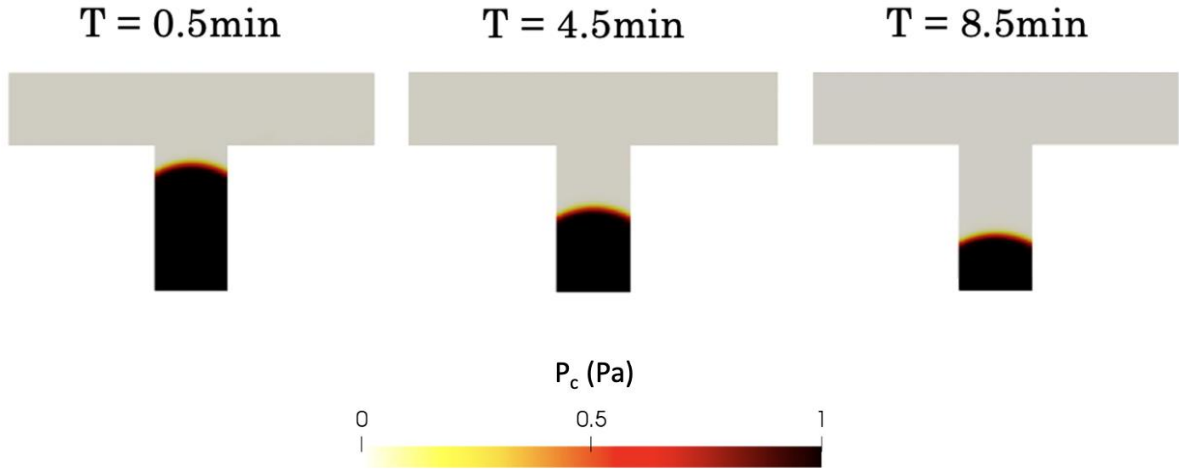

*Figure 8: Capilllary pressure for the SC geometry for times T =0.5, 4.5 and 8.5 minutes.*

6.2 Triple cavity configuration

Three experiments are conducted in three identical triple cavity (TC) micromodel geometries. In all three experiments at time $T = 0.41$ min water traps the $CO_2$ in the first cavity (TC_C1), at $T = 0.7$ min in the second cavity (TC_C2) and $T = 1$ min in the third cavity (TC_C3). The amount of $CO_2$ trapped in each cavity and the error ε between the amount of $CO_2$ trapped in each cavity between different experiments can be seen in Table 4. The experimental results show that the amount of $CO_2$ trapped inside the cavities is the same between the identical experiments with an error of 2%.

In Figure 9 and 11 we observe the initial trapping and the dissolution rate inside the three cavities. In all three experiments the amount of trapping as well as the dissolution of the

$CO_2$ bubbles over time are repeatable. Similar to SC case, the dissolution of the bubbles at the initial stage when it is close to the flow channel is quick, indicating an advective dissolution regime. As the bubbles dissolve and the interface moves further away from the channel, the dissolution rate decreases, which indicates a transition to a diffusive dissolution regime.

Table 4: Fraction of CO2 inside the cavities for the experiment conducted in the TC geometries.

| Geometry | Trapped $CO_2$ TC1 | Trapped $CO_2$ TC2 | Trapped $CO_2$ TC3 |
| --- | --- | --- | --- |
| TC_C1 | 0.90 | 0.89 | 0.88 |
| TC_C2 | 0.89 | 0.87 | 0.87 |
| TC_C3 | 0.88 | 0.89 | 0.89 |

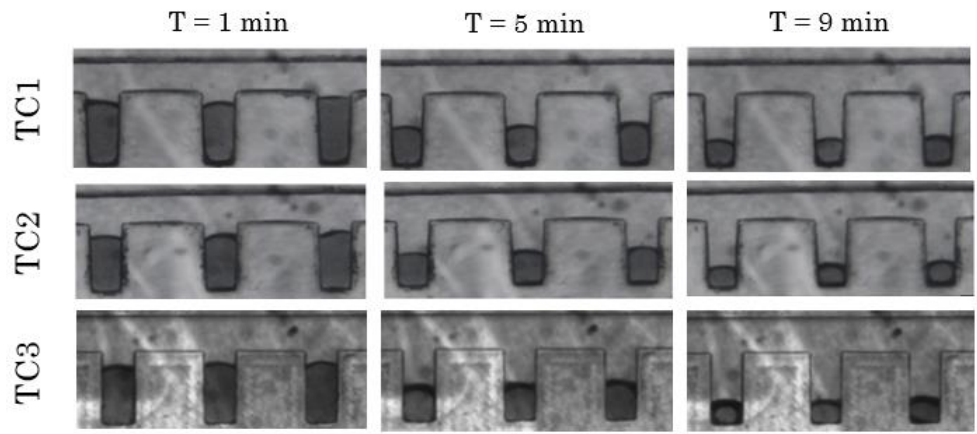

Figure 9: $CO_2$ bubble sizes observed during the experiment at the three identical TC geometries at times T = 1, 5 and 9 min.

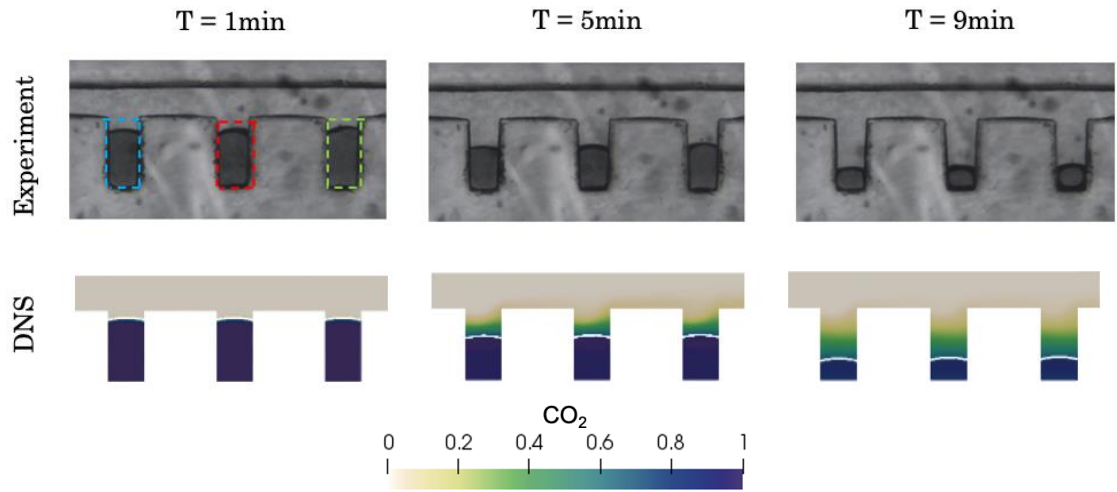

Figure 10: Dissolution of the $CO_2$ bubbles trapped in the triple cavity configuration. $CO_2$ bubble size observed during the experiment at times 1, 5, and 9 min (top). $CO_2$ bubble obtained from the DNS at times T= 1, 5 and 9 min (bottom). Cavities C1 (blue), C2 (red) and C3 (green).

In the DNS simulation, the bubbles are initialised inside the cavities so that the fraction of $CO_2$ in the middle plane matches the fraction of $CO_2$ trapped in the experiments conducted in the 3D printed micromodels. At $T = 0$ pure water is injected in the left side of the channel at a flow rate $Q = 1.67 \times 10^{-10}$ m$^3$.sec$^{-1}$ and Ca = $3.33 \times 10^{-6}$. The results of the numerical simulation can be seen in Figure 10 and Figure 11. When comparing the numerical simulation results to the experimental results, the dissolution of the bubble at times $T = 1, 5$ and $9$ min, the numerical simulation accurately captures the mass evolution of the $CO_2$ bubbles. In Figure 12, it can be also seen that the dissolution rate predicted by the numerical simulation matches the dissolution rate obtained from the experiments conducted in 3D printed micromodels. The overall error between the average experimental values and the direct numerical simulation for the three cavities is $\varepsilon_{C1} = 1.81\%$, $\varepsilon_{C2} = 1.86\%$ and $\varepsilon_{C3} = 2.52\%$. For the first (C1) and second cavity (C2) we can observe an increased deviation between the simulation and experimental results after 4 minutes and 6 minutes respectively. This deviation cannot be observed yet for the third cavity in the time frame of our experiment. For the third cavity (C3) the increased over all error between the numerical and experimental results ($\varepsilon_{C3} = 2.52\%$) could be due to experimental error or because during the experiment, when the water is injected in the geometry and traps the $CO_2$ in the first (C1) and second cavity (C2), the dissolution process has already started, and the water is partially saturated in $CO_2$ when it reaches the third cavity. This leads to a decrease in the concentration gradient between the water and the trapped $CO_2$ in the third cavity, which will have, as a result, a slower dissolution process as observed in the experiments. Initialising the bubbles inside the cavity has as a result to not take into account the aforementioned process introducing error to the overall system. Similar to the simulation results in SC, investigating this error further will require improving the computational efficiency of the solver.

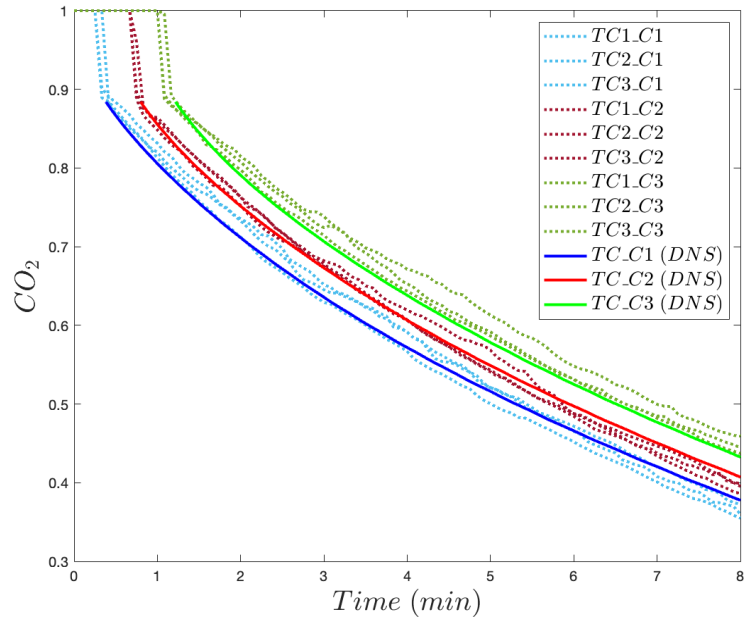

*Figure 11: Fraction of CO$_2$ in the middle plane for the three cavities C1(demonstrated in blue in Figure 11), C2 (demonstrated in red in Figure 11) and C3 (demonstrated in green in Figure 11) in the TC geometry as observed at the three identical experiments vs fraction of CO$_2$ predicted by the DNS..*

## 6.3 Multiple Channels

Three experiments are conducted in the MC geometry. The experimental results can be seen in Figure 12. In Figure 12 the experiments in the three micromodels are separated into three Stages. In Stage 1, the initial amount of $CO_2$ trapped inside the three identical micromodels when the injected water reaches the outlet can be seen. It can be observed that the amount of $CO_2$ is different inside the three identical micromodels which could possibly be due to artificial surface roughness being different in the three micromodel devices. In Stage 2, as the $CO_2$ dissolves inside the top channel and the five perpendicular channels, the only $CO_2$ in the geometries is located in the bottom channel. Finally, in Stage 3, it can be seen the invasion of the bottom channel in the three experiments due to the dissolution process. For the MC1 experiment the bottom channel is invaded by channel 4, for MC2 experiments it is invaded by channel 5 and for MC3 experiment by channel 1. As the trapped $CO_2$ phase dissolves, the water phase advances and invades the bottom channels. The invasion of the bottom channel from Stage 2 to Stage 3 strongly depends to the capillary entry pressure. The different pattern of invasion observed at the bottom channel in the three experiments also indicates that there is artificial roughness which impacts the flow displacement. Artificial surface roughness can lead to one of the perpendicular channels to have different size than the rest. This size difference can simultaneously impact the capillary entry pressure of the throat as well as the dissolution rate which leads to a preferential pathways during the invasion of the water in the geometry. In homogeneous cases like the MC geometry, where the five perpendicular channels have the same width, the displacement is strongly affected by small inconsistencies and artificial roughness. Moreover, at Stage 3, the saturation of $CO_2$ as well as the interface area is different for the three different experiments. The maximum amount of trapping is 89% (MC1) of the bottom channel while the minimum amount of trapping 84% (MC3) leading to an 6% variance between the experiments. Furthermore, the interfacial area for MC1 and MC3 is approximately $3mm^2$ while for MC2 is approximately $2.5\ mm^{2.}$ This lack of repeatability in the evolution of water does not allow to produce a benchmark experimental dataset in MC geometry which would allow comparison and validation of the numerical simulator.

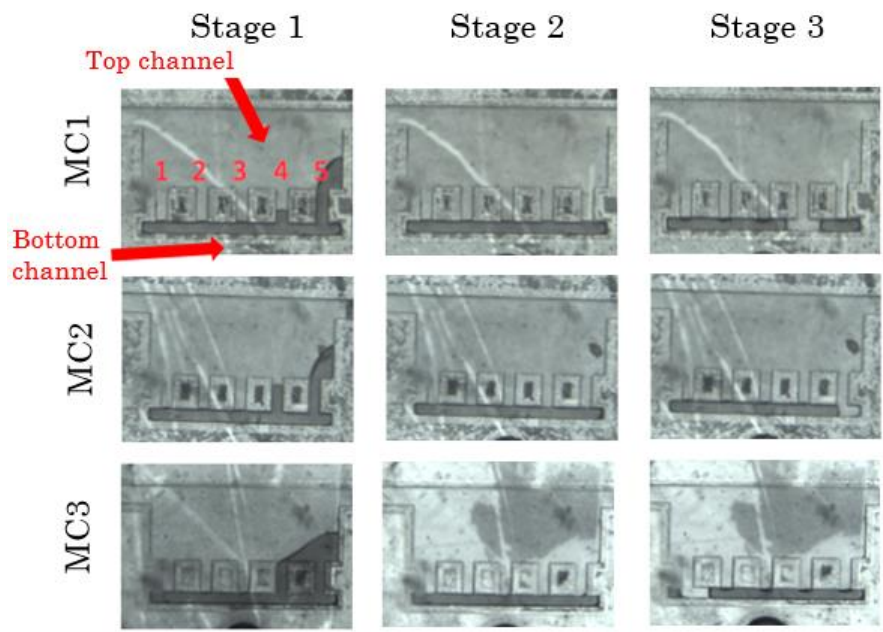

*Figure 12: $CO_2$ trapping at three identical multi-channel geometries. The amount of $CO_2$ trapped is different in each geometry.*

Furthermore, conducting a direct numerical simulation using a reduced surface tension is not possible. This method cannot be applied for the MC geometry since the dissolution is coupled with displacement of the $CO_2$. Using a different surface tension would significantly impact the simulation results. Finally, the mesh size is significantly larger and would lead to increased computational time.

## 6. Conclusions

In this work we use 3D printing technology to generate three micromodel geometries, the single cavity (SC), triple cavity (TC), and multiple channels geometry (MC). We print each geometry three times and conduct $CO_2$ trapping and dissolution experiments to assess the repeatability of experiments in each geometry. The experimental results show that in the cases of SC and the TC the trapping is repeatable for all three experiments with an error of 2%. Comparison of the temporal evolution of the $CO_2$ in three SC micromodels shows identical results. Moreover, comparison of the temporal evolution of the $CO_2$ inside the three TC geometries also shows reproducible results. In all experiments conducted in the SC and TC geometries, we initially observe a fast dissolution rate since the bubbles are close to the channel where water flow occurs. The flow strongly affects the bubbles dissolution leading to an advective dissolution regime. While the bubbles dissolves further away from the channel the dissolution rate decreases, and the dissolution is dominated by diffusion instead of advection. For the MC geometry, the initial trapping and dissolution pattern differs between different experiments. This could be due to the surface roughness affecting the trapping pattern. Surface roughness present in a homogeneous geometry, like the MC geometry, can have a strong impact in the flow pattern progression. Thus, the experimental results obtained for the MC geometry cannot be used as a benchmark dataset. While previous works has utilised micromodel devices to investigate trapping and dissolution of $CO_2$ [25, 26], this is the first experimental work that provides repeatable results where $CO_2$ bubbles are trapped and dissolved under known boundary conditions. The known boundary conditions during the experiment allows for direct comparison with DNS.

The capillary number for experiments conducted in the SC and TC geometries is $Ca = 3.33 \times 10^{-6}$. The low capillary number doesn't allow for performing accurate direct numerical simulation due to spurious currents. Therefore, to simulate the phenomenon for the single and triple cavity configurations, we initialise the bubble inside the cavity and simulate only the dissolution process of the experiment using a surface tension of $\sigma/100$. This doesn't allow to simulate the trapping of the bubbles since the surface tension strongly impacts the displacement of the fluids, but it allows for calculation of the dissolution of $CO_2$ inside the SC and TC geometry by initializing the bubbles inside the cavity. For the SC geometry DNS accurately captures the dissolution and matches the experimental results obtained from the 3D printed micromodels. For the TC model, the DNS also accurately captures the dissolution for the first cavity (C1) and the second cavity (C2). We observe a small deviation manifesting for the third cavity that could be due to initialising the bubbles inside the cavity, which introduces error to the overall system.

Development of the numerical models to capture capillary dominated processes for multiphase flow that overcome existing problems like capillary waves and spurious currents is required [51]. With this work, we produce an experimental dataset that allows for validation of numerical models that capture multiphase flow displacement with interfacial transfer of a gas to an aqueous phase. Finally, we show that although simple scenarios like dissolution of $CO_2$ in a simple cavity can be accurately captured and we highlight the need for faster computational methods capable of solving low capillary number scenarios in more complicated geometries to be able to accurately calculate $CO_2$ dissolution in a complicated rock pore-space scenario.

## Declaration of Competing Interest

The authors declare that they have no known competing financial interests or personal relationships that could have appeared to influence the work reported in this paper.

## Data availability

The experimental dataset can be found at the Open Science Framework (OSF) repository https://osf.io/zynxv/.

## Acknowledgements

This work was funded by Engineering and Physical Sciences Research Council's (EPSRC) grant on Direct Numerical Simulation for Additive Manufacturing. (EP/P031307/1).

SR and MMB thank the Agence Nationale de la Recherche (ANR) under the project ANR-21-CE50-0038-01 for partially funding this research.